\documentclass[twocolumn,aps,superscriptaddress,multicol,amsmath,amssymb]{revtex4-2}
\usepackage{amssymb}
\usepackage{amsmath}
\usepackage{graphicx}
\usepackage{epstopdf}
\usepackage{bm}
\usepackage{gensymb}
\usepackage{soul}
\usepackage{sidecap}
\usepackage[normalem]{ulem}
\usepackage{times}
\usepackage{mathrsfs}

\usepackage{float}
\usepackage[caption = false]{subfig}
\usepackage{subfig}
\usepackage{enumerate}
\usepackage{multirow}
\usepackage{tabularx}
\usepackage{array}
\usepackage{url}
\usepackage{slantsc}
\usepackage{lmodern}
\newcommand\redout{\bgroup\markoverwith{\textcolor{red}{\rule[.5ex]{2pt}{0.4pt}}}\ULon}

\usepackage[colorlinks=true,citecolor=blue]{hyperref}
\hypersetup{colorlinks=true,citecolor=blue,linkcolor=red,urlcolor=blue}

\usepackage{hyperref,cleveref}
\newcommand{\be}{\begin{equation}}
\newcommand{\ee}{\end{equation}}
\newcommand{\bk}{{{\bf{k}}}}
\newcommand{\bq}{{{\bf{q}}}}

\newcommand{\bg}{{{\bf{g}}}}

\newcommand{\bea}{\begin{eqnarray}}
\newcommand{\eea}{\end{eqnarray}}

\newcommand{\bd}{\begin{displaymath}}
\newcommand{\ed}{\end{displaymath}}
\newcommand{\ba}{\begin{array}}
\newcommand{\ea}{\end{array}}
\newcommand{\bi}{\begin{itemize}}
\newcommand{\ei}{\end{itemize}}
\newcommand{\bc}{\begin{center}}
\newcommand{\ec}{\end{center}}
\newcommand{\bfl}{\begin{flushleft}}
\newcommand{\efl}{\end{flushleft}}
\newcommand{\bfr}{\begin{flushright}}
\newcommand{\efr}{\end{flushright}}
\newcommand{\no}{\nonumber}

\newcommand{\mi}{\rm i}

\newcommand{\bl}{\begin{aligned}}
\newcommand{\el}{\end{aligned}}

\def\bk{{\bf k}} \def\bq{{\bf q}} \def\bp{{\bf p}} 
\def\bg{{\bf g}}  \def\bd{{\bf d}}

\def\6{\partial}

\def\={\!\!\!&=&\!\!\!}
\def\+{\!\!\!&&\!\!\!+~}
\def\-{\!\!\!&&\!\!\!-~}

\graphicspath{{.}{./Figs/}}
\usepackage{color}
\usepackage[dvipsnames]{xcolor}
\usepackage{xcolor}
\begin{document}
\title{
Entangled Photon Generation through Cooper Pair Recombination in a Noncentrosymmetric Quantum Well
}
\author{Mehdi Biderang}
\affiliation{Department of Physics, University of Toronto, 60 St. George Street, Toronto, Ontario, M5S 1A7, Canada}
\affiliation{DelQuanTech Inc., 1130-500 Doris Ave., Toronto, Ontario, M2N 0C1, Canada}
\affiliation{Asia Pacific Center for Theoretical Physics (APCTP), Pohang, Gyeongbuk, 790-784, Korea}
\author{Erfan Hosseini}
\affiliation{Institute for Quantum Computing, University of Waterloo, 200 University Ave W, Waterloo, Ontario, N2L 3G1, Canada}
\author{Alireza Akbari}
\affiliation{Beijing Institute of Mathematical Sciences and Applications (BIMSA), Huairou District, Beijing, 101408, P. R. China}
\affiliation{Institut für Theoretische Physik III, Ruhr-Universität Bochum, D-44801 Bochum, Germany}
\affiliation{Asia Pacific Center for Theoretical Physics (APCTP), Pohang, Gyeongbuk, 790-784, Korea}
\date{\today}

\begin{abstract}
We explore theoretically the generation of entangled two-photon pairs by Cooper pair recombination in a noncentrosymmetric [001]-quantum well superconductor, driven by a forward-biased p-n junction with a superconducting layer which exhibits admixture Rashba and Dresselhaus spin-orbit couplings.
We show that the highest achievable purity of entangled photon pairs emerges within scenarios involving pure singlet  Cooper pairs,
specifically, the conventional $s$-wave gap function.
Our results highlight the importance of minimizing the charge-carrier level concentration and balancing the magnitudes of Rashba and Dresselhaus spin-orbit couplings to achieve entangled states with enhanced purity,
which can be realized by reducing the amplitudes of antisymmetric spin-orbit couplings.
In addition to purity concerns, to  explore the distribution of two-photon states, we compare their population across entangled pairs 
for potential superconducting pairings.
\end{abstract}

\maketitle
\section{Introduction}
Entanglement  
stands as a cornerstone in modern quantum technologies due to its   significant role in a spectrum of applications, notably in quantum cryptography~\cite{Bennet_PRL_1993,Bouwmeester_Nature_1997,Gisin_Rev_Mod_Phys_2002,Yin_Nature_2020,Schimpf_SciAdv_2021,Pseiner_IOP_2021}, computing~\cite{Raussendorf_PRL_2001,Obrien_Science_2007,Laad_Nature_2010,Eldredge_PRR_2020,Conlon_Nature_2023}, communication~\cite{Hu_PRL_2021,Zou_IOP_2021,Piveteau_NatComm_2022} and metrology~\cite{Mitchell_Nature_2004,Polino_AVSQuantum_2020,Long_PRL_2022,Colombo_Nature_2022}. 
Achieving reliable quantum entanglement sources can make big advancement in classical sensing tools like radars and lidars, which are using entangled photons to enhance the detection capability of the system over the classical limit~\cite{Tan_PRL_2008,Degen_RevModPhys_2017,Pirandola_Nature_2018,Barzanjeh_arXiv_2023}.
The quest for robust sources capable of producing entangled photon pairs has become paramount for advancing these applications. In this pursuit, the convergence of semiconductor and superconductor technologies has birthed an interdisciplinary domain termed superconducting optoelectronics~\cite{Shainline_JAPpPhys_2019,Shaneline_PRAPP_2016,Khan_NatElec_2022}. 
This field has witnessed significant strides with the creation of hybrid devices, including but not limited to, superconducting light-emitting diodes, quantum dots~\cite{Mou_IEEE_2015,Mou_PRB_2015,Sabag_PRB_2017}, and superconductor-based waveguide amplifiers~\cite{Andrews_IEEE_2013,Marjieh_NewJourPhys_2016,Qasymeh_Nature_2022}.

The complex interdependencies between superconductivity, semiconducting heterostructures, and associated quantum effects form a rich foundation for further exploration. This confluence opens avenues for advancing both fundamental understanding and practical applications in the realm of quantum information processing~\cite{Benito_APL_2020,Burkard_Nature_2020,Yuan_Vac_2021}.
The emergence of Rashba and Dresselhaus spin-orbit couplings (SOCs) in zinc-blende structures and quantum wells stands as a distinctive feature that significantly shapes the electronic properties of these systems~\cite{Dresselhaus_PR_1955,Ganichev_PSS_2014}. In noncentrosymmetric crystals like zinc-blende semiconductors, the lack of inversion symmetry gives rise to the Rashba SOC, an effect resulting from structural asymmetry. The Rashba term induces a momentum-dependent splitting of electronic bands, impacting charge carriers' spin dynamics. In the presence of an asymmetric potential gradient, such as that encountered at interfaces or within quantum wells, the Dresselhaus SOC becomes prominent. Originating from structural asymmetry perpendicular to the interface, the Dresselhaus term contributes an additional layer to the spin-orbit interaction. Both Rashba and Dresselhaus SOCs play a crucial role in the manipulation and control of spin states, paving the way for novel electronic and optoelectronic phenomena in these materials~\cite{Koralek_Nature_2009,Balocchi_PRL_2011,Walser_Nature_2012,Kawano_Nature_2019,Lu_IOP_2020}.

The creation of a forward-biased p-n junction with a superconducting layer (P-N-S) heterostructure is an intriguing path that has been suggested for the production of entangled photons. At this point the proximity effect may be able to induce superconductivity in the n-type semiconductor~\cite{Gordon_PRB_2018}. 
Furthermore, the strategic integration of a quantum well structure within the semiconductor layers not only facilitates the production of entangled photon pairs but also enables the generation of pure polarization-entangled photons~\cite{Hayat_PRB_2014}. These design intricacies exploit the Rashba and Dresselhaus SOCs, which play a pivotal role in manipulating the parity of the superconducting order parameter~\cite{Yip_PRB_2002,Edelstein_PRB_2003}.

Recent investigations have considered the manifestations of Cooper-pair-based two-photon gain in semiconductor-superconductor structures. These studies have unveiled a broadband enhancement in ultrafast two-photon amplification, delineating a comprehensive quantum-optical model encompassing both singly- and fully-stimulated two-photon emission~\cite{Marjieh_NewJourPhys_2016}. Moreover, the exploitation of inherent angular momentum entanglement within the superconducting state has showcased the generation of polarization-entangled photons through Cooper-pair luminescence, circumventing the necessity for isolated emitters within semiconductors~\cite{Hayat_PRB_2014}.
Further explorations into the electroluminescence and photonic attributes of forward-biased p-n junctions in the proximity of superconducting media have revealed notable enhancements in electroluminescence within specific frequency ranges in the presence of superconductivity~\cite{Hlobil_PRB_2015}.  The strategic coupling of superconducting contacts with quantum dots has not only yielded intensified luminescence at temperatures below  superconducting critical temperature, $T^{}_{c}$, but has also yielded higher purity in photon entanglement, mitigating the adverse effects of excitonic energy level splittings~\cite{Bouscher_JourOptics_2017}.
Moreover, investigations into the purity of generated photons have extended to triplet superconductors by examining a fixed spin-triplet pairing orientation, 
uncovering intriguing directional dependencies.
Specifically, the induction of triplet pairing via Rashba spin-orbit coupling  has been identified as a mechanism for generating pure entangled photons when the photon polarization axis aligns parallelly with spin-triplet pairing orientation. 
Conversely, induced triplet pairing within a singlet superconductor has exhibited a degradation in state purity. Strikingly, induced singlet pairing in a triplet superconductor has shown an amplification in the production of entangled pairs~\cite{Gordon_PRB_2018}. These findings underscore the intricate interplay between superconductivity, spin-orbit interactions, and resultant entanglement properties within these heterostructures.

Here, we study the effect of combination of Rashba and Dresselhaus SOCs and varying the relative amplitude of singlet to triplet superconductivity on the entanglement properties of the generated pairs of photons during the process of Cooper pair recombination within a semiconductor-superconductor heterostructure.
%

\section{Theoretical Model} 
\label{Sec:Model}

\begin{figure}[t]
	\includegraphics[width=0.99\linewidth]{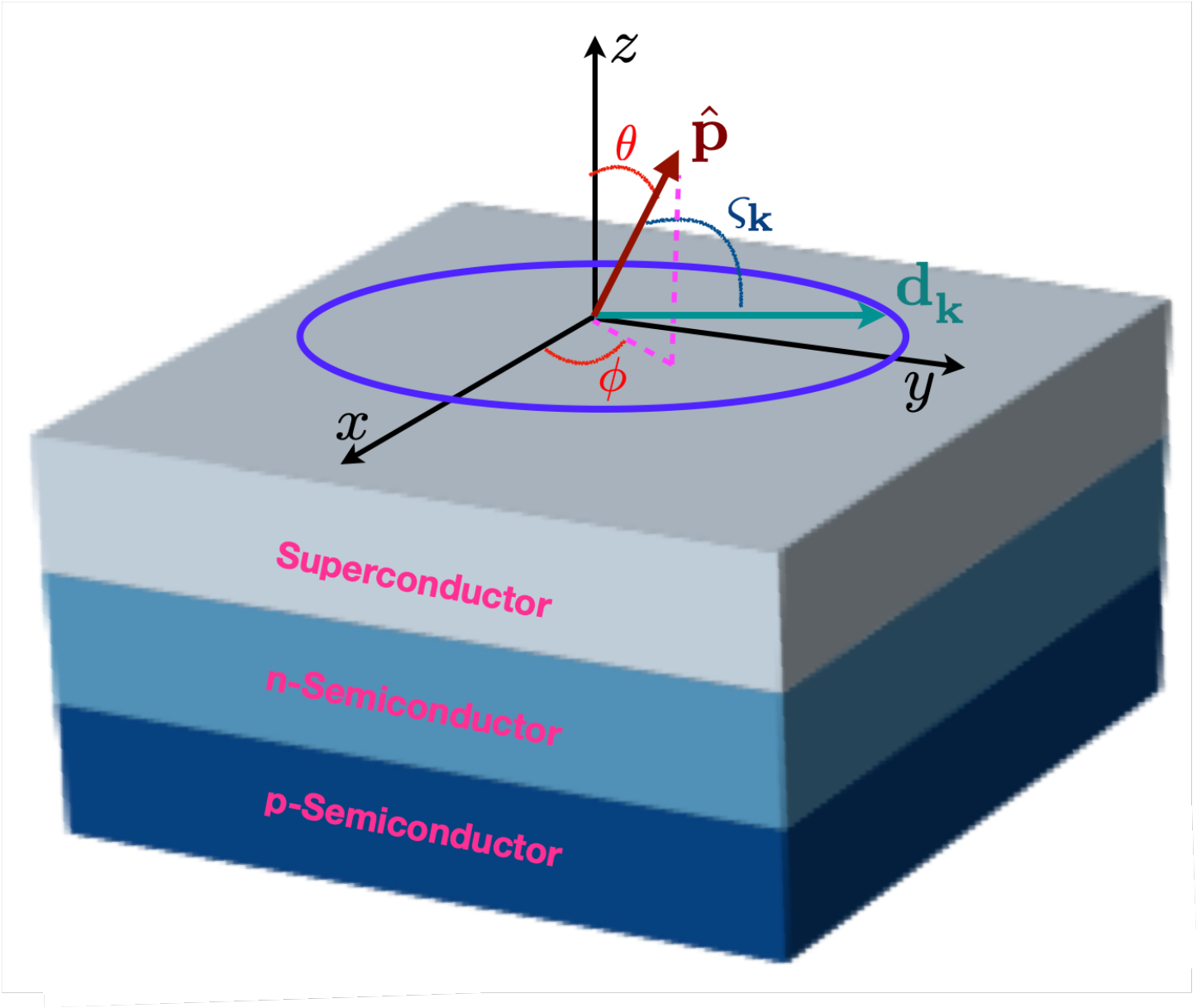}
	\caption{
            The schematic geometry of a semiconductor-superconductor heterostructure (P-N-S), where superconductivity is induced within the n-type semiconducting layer owing to the proximity of the s-wave superconductor.
		Vectors $\hat{\bp}$ and  $\bd^{}_{\bk}$
            symbolize the photon's polarization axis, and the direction of spin-triplet pairing, respectively, which have an  angle $\varsigma^{}_{\bk}$. 
	}
	\label{Fig:Fig1}
\end{figure}

The system under investigation is a P-N-S heterostructure, a configuration expounded in the approach outlined in Ref.~\cite{Gordon_PRB_2018}.
As depicted in Fig.~\ref{Fig:Fig1}, this composite structure merges a superconductor with a semiconducting medium (p-n junction). 
Within this arrangement, superconductivity manifests in the  semiconducting region (n-type) through the infusion of Cooper pairs from the superconducting contacts, facilitated by the proximity effect.
The upper valence bands of a p-n junction of the zinc-blende type semiconductor consist of heavy-hole (HH) and light-hole (LH) bands, which are degenerated at the zero crystal momentum. In a quantum well structure, as a result of crystal field, the degeneracy between these bands is removed, and this separation plays a key role in generating pure entangled photons through the recombination of Cooper pairs with the LH band~\cite{Hayat_PRB_2014}.
%

%
%
%

\subsection{Model  Hamiltonian}
We start our investigation by considering the intrinsic physics of a p-n junction,  which are described by the following Hamiltonian~\cite{Gordon_PRB_2018,Hayat_PRB_2014}
%
\begin{equation}
\bl
	{\cal H}^{}_{0}
 &
 =
	{\cal H}^{}_{\rm ph}
	+
	{\cal H}^{}_{\rm v}
	+
	{\cal H}^{}_{\rm c}
 \\
 &
 =
        \sum_{\bq,p}^{}
        \omega^{}_{\bq} a^{\dagger}_{\bq,p}a^{}_{\bq,p}
        +
        \sum_{\bk,J}^{}\varepsilon^{}_{\bk} h^{\dagger}_{\bk,J}h^{}_{\bk,J}
        +
        \sum_{\bk,j}^{}\xi_{\bk} c^{\dagger}_{\bk,j}c^{}_{\bk,j}
        ,
 \el
\end{equation}
%
%
where 
${\cal H}^{}_{\rm ph}$,
${\cal H}^{}_{\rm v}$, 
and
${\cal H}^{}_{\rm c}$
represent the Hamiltonian of photons, valence 
 (heavy-holes), and conduction electrons, respectively.
In this context, $a^{\dagger}_{\bq,p}$ represents the creation operator for a phonon with momentum $\bq$ and polarization $p=\pm 1$.
Additionally, the operators  $c^{\dagger}_{\bk j}$ ($c^{}_{\bk j}$)  and
$h^{\dagger}_{\bk,J}$  
 ($h^{}_{\bk,J}$) create (annihilate)    an electron and a hole with momentum $\bk$, respectively, located in the  conduction and heavy-hole  bands.
The angular momentum of electrons and holes is denoted by $j=\pm1/2$, and $J=\pm3/2$, respectively.
The process of electron-hole recombination and the generation of photons are described by the light-matter interaction within the dipole approximation, which can be expressed as:
%
%
\begin{equation}
\bl
	{\cal H}^{}_{\rm int}=
	\sum^{}_{\bk,\bq}
    \sum_{J=\mp \frac{3}{2}; p=\pm 1}^{}
	{\mathbb B}^{}_{\bk,\bq} a^{\dagger}_{\bq,p}
	h^{}_{\bq-\bk,-J}
	c^{}_{\bk,J+p}
	+ 
	{\rm H.c.},
 \el
	\label{Eq:potential}%
\end{equation}
%
%
where, ${\mathbb B}^{}_{\bk,\bq}$ denotes the scattering  matrix elements governing the process between the electron and hole states with absorbing/emitting photon with momentum $\bq$.
The summation over $p$ and $J$ fulfils the conservation of angular momentum during processes such as electron-hole recombination or pair creation requires that $p=J+j$.
\\

%
\subsection{Anisotropic Spin-Orbit Coupling}
\begin{figure}[t]
\includegraphics[width=1.0\linewidth]{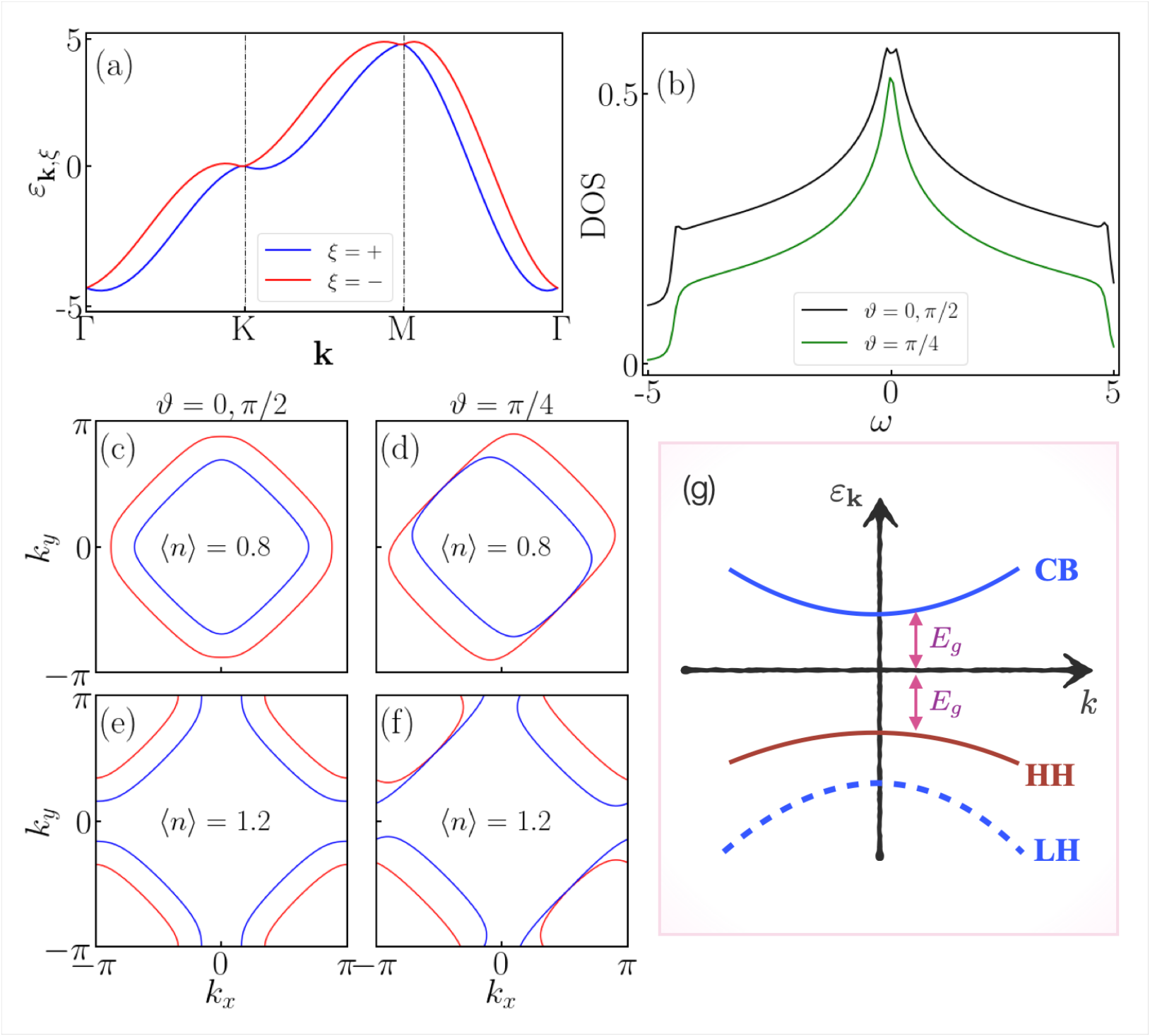}
\caption{ 
	(a) Energy dispersion of helical bands along the $\Gamma$KM$\Gamma$ path for pure Rashba/Dresselhaus SOC.
	(b) Density of states (DOS) for conduction electrons: The black curve depicts the pure Rashba ($\vartheta=0$) and pure Dresselhaus ($\vartheta=\pi/2$) cases, while the green curve represents equal contributions of Rashba and Dresselhaus effects ($\vartheta=\pi/4$). 
    For a clearer illustration, the black curve is shifted by 0.1.
	(c) and (d) represent the Fermi surface topology at $\langle n \rangle=0.8$ for $\vartheta=0$ and $\vartheta=\pi/2$, respectively.
	(e) and (f) Similar visualizations for a filling of $\langle n \rangle=1.2$.
     (g) Schematic diagram of 
    band structure in a semiconducting quantum well indicating the gap ($\ge 2E_g$) between electron states in the conduction band (CB) and their particle-hole symmetric counterpart in the heavy-hole (HH) states.
    The light-hole (LH) band is shown to complete the picture according to a typical $J=3/2$ system. 
	}
\label{Fig:Fig2}
\end{figure}

The effective Hamiltonian, describing the behavior of conduction electrons due to surface-induced asymmetry, is determined using an anisotropic spin-orbit coupled single-band tight-binding model that considers interactions between nearest neighbors.
As a result, the expression for conduction electrons in a two-dimensional (2D) square lattice structure, which lacks a center of inversion, is defined by~\cite{Frigeri_2006,Fujimoto_2007}
%
%
\begin{equation}
{\cal H}^{}_{\rm c}(\vartheta)=
\sum^{}_{k,jj'}
\Big[
\epsilon^{}_{\bk}
\hat{\sigma}^{}_{0}
+
\lambda
\bg^{}_{\bk}(\vartheta)
\cdot
\hat{\boldsymbol \sigma}
\Big]^{}_{jj'}
c^{\dagger}_{\bk j}
c^{}_{\bk j'}
+{\rm H.c.},
\label{Eq:Ham_rRashba_Dresselhaus}
\end{equation}
%
%
%
Here, $j=1/2$ ($j=-1/2$) refers to an electron state 
$\mid \uparrow\rangle$ 
($\mid \downarrow\rangle$).
%
The term $\epsilon^{}_{\bk}=-\mu-2t(\cos k_x+\cos k_y)$ represents the kinetic energy, with $\mu$ as the chemical potential and $t$ as the amplitude for first-neighbor hopping.
Moreover, $\hat{\sigma}^{}_{0}$ denotes the $2\times 2$ identity matrix, while $\hat{\sigma}^{}_{i}$ ($i=x,y,z$) represents the $i^{\rm th}$ component of the Pauli matrices in the spin space.
The antisymmetric vector 
%
\begin{equation}
\bg^{}_{\bk}(\vartheta)
=
\alpha(\vartheta) \bg^{\rm R}_{\bk}+\beta(\vartheta) \bg^{\rm D}_{\bk}   
\end{equation}
%
characterizes the strength  of the spin-orbit coupling, $\lambda$, combining the Rashba and Dresselhaus effects.
The dimensionless coefficients 
$\alpha(\vartheta)=\cos\vartheta$ 
and
$\beta(\vartheta)=\sin\vartheta$ 
indicate the magnitudes of the Rashba and Dresselhaus SOCs, respectively. The parameter $\vartheta\!\in\![0,\pi/2]$ modulates their relative contributions.
In our model, the Rashba and Dresselhaus $\bg$-vectors are given by,
%
\begin{equation}
        \bg^{\rm R}_{\bk}=(\sin k_y,-\sin k_x);
        \quad 
        \bg^{\rm D}_{\bk}=(\sin k_x,-\sin k_y).
\end{equation}
%
%
In our calculations, we perform a projection of the conduction band Hamiltonian onto an effective pseudospin-1/2 model through the transformation
${\cal P}^{}_{\pm}{\cal U}{\cal H}^{\prime}_{\rm c}{\cal U}{\cal P}^{}_{\pm}$.
Here, ${\cal P}^{}_{\pm}$ projects the conduction band onto the helical
bands with $\xi=\pm 1$, while ${\cal U}$ diagonalizes the Hamiltonian of the conduction electrons.
The energy dispersion in the presence of SOC is defined by
%
\begin{equation}
	\varepsilon^{}_{\bk,\xi}(\vartheta) = 
	\epsilon^{}_{\bk}
	+\xi\lambda\sqrt{
		\sin^2\!k_x+\sin^2\!k_y
		\!+2
		\sin 2\vartheta
		\sin\!k_x \sin\!k_y
	},
	\label{Eq:Normal_Energy}
\end{equation}
%
when $\vartheta=0$ ($\pi/2$), corresponding to the pure Rashba (Dresselhaus) case, the energy dispersion simplifies to 
%
%
\begin{equation}
     \no
    \varepsilon^{}_{\bk,\xi}
    (\vartheta=
    0,\pi/2)
    =
    \epsilon^{}_{\bk}+\xi\lambda\sqrt{
    	\sin^2\!k_x+\sin^2\!k_y
    },
\end{equation}
%
%
indicating a complete $C^{}_{4v}$ point group symmetry of the crystal lattice.
We show the corresponding  energy dispersion of these helical bands along the
$\Gamma{\rm  KM}\Gamma$ path in  Fig.~\ref{Fig:Fig2}(a).
However, the energy dispersion is expressed as 
%
\begin{equation}
     \no
\varepsilon^{}_{\bk,\xi}(\vartheta=\pi/4)
=
\epsilon^{}_{\bk}+\xi\lambda
|\sin k_x+\sin k_y|
\end{equation}
%
%
at $\vartheta=\pi/4$, where Rashba and Dresselhaus SOCs contribute equally.
Under this condition, the Hamiltonian symmetry reduces to $C^{}_{2v}$, leading to helical Fermi surfaces intersecting at $k_y\!=\!-k_x$.
%
%
In Fig.~\ref{Fig:Fig2}(b), the density of states (DOS) of the noncentrosymmetric n-type semiconductor is presented 
for three different scenarios: $\vartheta=0$ represents the pure Rashba case, $\vartheta=\pi/4$ corresponds to equal contributions of Rashba-Dresselhaus, and $\vartheta=\pi/2$ signifies the pure Dresselhaus spin-orbit coupling.
The Fermi surface (FS) topology is depicted in 
Figs.~\ref{Fig:Fig2}(c)~and~\ref{Fig:Fig2}(e) for pure Rashba or pure Dresselhaus SOC at filling fractions 
$\langle n \rangle=0.8$
(hole-doping) and 
$\langle n \rangle=1.2$ 
(electron-doping), respectively.
Note: Half-filling with one electron per state indicates $\langle n \rangle > 1$ for electron doping and $\langle n \rangle < 1$ for hole doping.
A comparison with the DOS reveals the occurrence of van Hove singularities (Lifshitz transitions) at half-filling, leading to changes in the FS topology.
Lastly, Figs.~\ref{Fig:Fig2}(d)~and~\ref{Fig:Fig2}(f) exhibit the FS texture for the aforementioned filling levels in the presence of equal contributions of Rashba and Dresselhaus SOCs.
The prominent consequence of the antisymmetric SOC is the lifting of the two-fold spin degeneracy and the formation of helical bands characterized by helicity $\xi=\pm 1$.
\\

\subsection{Superconducting state}
Our objective is to investigate the impact of different SOCs contributions at specified filling fractions and various types of superconducting gap symmetries. This exploration aims to assess the purity of the generated entangled photons resulting from the breakdown of Cooper pairs.
The broken inversion symmetry within the crystal structure results in the violation of parity.
Consequently, in  the ground state of superconducting regime,
\be\bl
{\cal H}_{\rm SC}=
\sum_{\mathbf{k},jj'} 
\Big( \hat{\Delta}^{}_{jj'}(\bk) c_{\mathbf{k},j}^\dagger c_{-\mathbf{k},j'}^\dagger + 
{\rm H.c.}
\Big),
\el\ee
 both spin-singlet and spin-triplet pairings will simultaneously condense.
It is widely acknowledged that within the spin space, the superconducting gap function $\hat{\Delta}^{}_{}(\bk)$ can be represented by a $2\times 2$ matrix, formulated as follows:
%
\begin{equation}
\hat{\Delta}^{}_{}(\bk)=
\hat{\Delta}^{\rm sing}_{k}
+
\hat{\Delta}^{\rm trip}_{k}=
{\mi}
\Big[
\psi^{}_{\bk}\hat\sigma^{}_{0}
+ 
\bd^{}_{\bk}\cdot \hat{\boldsymbol{\sigma}}
\Big]
\hat{\sigma}^{}_{y}.
\label{Eq:deltamixed}%
\end{equation}
%
%
In this expression, $\psi^{}_{\bk}$ and $\bd^{}_{\bk}$ denote the momentum-dependent spatial components of the spin-singlet and spin-triplet constituents of the superconducting order parameters, respectively~\cite{sigrist2009}. 
These components are characterized as even and odd functions with respect to the momentum $\bk$~\cite{Samokhin_2004}.
The projection of the superconducting gap function onto the helical bands gives rise to $\Delta^{}_{\bk,\xi}=\psi^{}_{\bk}+\xi|\bd^{}_{\bk}|$.
The translational invariant structure factors of the permissible irreducible representations of the 
$C^{}_{4v}$ point group within a square lattice can be found in the Table~\ref{Tab:Stracture_Factor}.
%
\begin{table}[t]
	\begin{center}
		\caption{Structure factors ($f^{}_{\bk}$) of the singlet and triplet superconducting pairings for the allowed irreducible representations of the $C^{}_{4v}$ point group in the presence of antisymmetric SOC.}
		\label{Tab:Stracture_Factor}
		\addtolength{\tabcolsep}{9pt}
		\begin{tabular}{c c c}
			\hline
			 $\ell$ & Symmetry &  $f^{}_{\bk}$  \\
			\hline
			 0 & $s$-wave & $1$ \\0 & $s^*$-wave & $\cos k_x+\cos k_y$ \\ 
            1 & $p_x$ ($p_y$)-wave & $\sin k_x \;(\sin k_y)$ \\ 
			2 & $d^{}_{xy}$-wave & $\sin k_x \sin k_y$  \\ 
			2 & $d^{}_{x^2-y^2}$-wave &  $\cos k_x-\cos k_y$ \\
			3 & $f^{}_{x(x^2-y^2)}$-wave &  $\sin k_x(\cos k_x-\cos k_y)$ \\	
			\hline
		\end{tabular}
	\end{center}
\end{table}
%
We posit that in the triplet channel, electrons are exclusively condensed in $p$-wave pairing, but we also examine the $f$-wave pairing. Therefore, the potential scenarios for the superconducting ground state  encompass $s+p$, $s^{*}+p$, $d^{}_{x^2-y^2}+p$, and  $d^{}_{xy}+p$, as well as,
$s+f$
pairings.
By introducing the dimensionless coefficient 
$0<r<1$
to quantify the contribution of even and odd parity pairings on the Fermi surface, the superconducting gap functions in the singlet and triplet channels are expressed as 
\be
\bl
\psi^{}_{\bk}
= 
r \Delta^{}_{0} f^{}_{\bk};
\quad
\bd^{}_{\bk}
=
(1-r)\Delta^{}_{0}{\bg}^{}_{\bk}.
\el
\ee
It is widely recognized that the stability of triplet superconductivity is contingent on $\bd^{}_{\bk}||\bg^{}_{\bk}$~\cite{sigrist2009}.
Here, $\Delta^{}_{0}$ represents the magnitude of the superconducting order parameter.
The structure factor $f^{}_{\bk}$ encapsulates the orbital angular momentum of pairing within the superconducting gap, as elucidated in Ref.~\cite{Biderang_PRB_2018}.
\\

\subsection{Random phase approximation consideration}
Given our focus on spin fluctuations as a potential mechanism for electron pairing and superconductivity, our investigation commences with an examination of spin susceptibility. 
We will analyze its dependence on factors such as next-nearest-neighbor hopping, spin-orbit coupling, and doping.
The Matsubara Green's function for a system of non-interacting electrons with antisymmetric SOC (ASOC) is represented by 
\be\bl
\hat{\cal G}^{0}_{}(k,\mi\omega_n)=
[(\mi\omega_n-\epsilon^{}_{\bk})\hat{1}-\lambda \bg^{}_{\bk}\cdot \hat{\boldsymbol{\sigma}}]^{-1}.
\el
\ee
Here, $\omega_n = (2n + 1)\pi T$ and $\hat{1}$ denote the fermionic Matsubara frequencies and a $2\times 2$ identity matrix, respectively.
Performing a unitary transformation, the spin and band space Matsubara Green's functions are related together by
%
\begin{equation}
\bl
	\hat{\cal G}^{0}_{}(\bk,\mi\omega_n)=
	\frac{1}{2}\sum_{\xi=\pm 1}
	[\hat{1}+\xi \hat{\bg}^{}_{\bk}\cdot \hat{\boldsymbol\sigma}]
	{\cal G}^{0}_{\xi}(\bk,\mi\omega_n),
	\label{Eq:Green_0_Spin_Band}%
 \el
\end{equation}
%
where $\hat{\bg}^{}_{\bk}={\bg}^{}_{\bk}/|{\bg}^{}_{\bk}|$, and ${\cal G}^{0}_{\xi}(\bk,\mi\omega_n)=[\mi\omega_n-\varepsilon^{}_{\bk,\xi}]^{-1}$ shows the free electron Matsubara Green’s function in band space.
In order to study the spin fluctuations and electron instabilities within the system, we opt for the conventional random phase
approximation (RPA) methodology.
Within the context of linear response theory, the tensor of bare spin susceptibility, 
at the momentum $\bq$ and bosonic Matsubara frequencies $\omega_m = 2m\pi T$, is defined as
%
\begin{equation}
 \bl
	\Big[\hat{\chi}^{(0)}_{}(\bq,\mi\omega_m)\Big]^{\sigma_3\sigma_4}_{\sigma_1\sigma_2}=&
 \\
 &\hspace{-1cm}
	-\frac{T}{4N}
	\sum_{\bk,\mi\nu_n}
	\hat{\cal G}^{0}_{}(\bk,\mi\nu_n)
	\hat{\cal G}^{0}_{}(\bk+\bq,\mi\omega_m+\mi\nu_n).
	\label{Eq:Bare_Kappa}%
 \el
\end{equation}
%
Performing summation over the fermionic Matsubara frequencies $\mi\nu_n$ and an analytical continuation $\mi\omega_m\rightarrow \omega+\mi 0^+$, the retarded bare susceptibility is given by the following Lindhard function
%
\begin{equation}
 \bl
	\Big[\hat{\chi}^{(0)}_{{\rm Ret}}(\bq,\omega)\Big]^{\sigma_3\sigma_4}_{\sigma_1\sigma_2}
 =&
\\
&\hspace{-1cm}
\frac{1}{4N}
	\!\sum_{\bk,\mi\nu_n}
W^{\bk\bq,\xi\xi'}_{\sigma^{}_1\sigma^{}_2\sigma^{}_3\sigma^{}_4}
	\frac{n_F(\varepsilon^{}_{\bk+\bq,\xi'})-n_F(\varepsilon^{}_{\bk,\xi})}{\omega+\varepsilon^{}_{\bk,\xi}-\varepsilon^{}_{\bk+\bq,\xi'}+\mi 0^+},
	\label{Eq:Bare_Kappa_Lindhard}%
 \el
\end{equation}
%
where $n^{}_{F}(\varepsilon)=[1+\exp(\varepsilon/T)]^{-1}$ represents the Fermi-Dirac distribution function of electrons at temperature $T$.
Moreover, the weight factor $W^{\bk\bq,\xi\xi'}_{\sigma^{}_1\sigma^{}_2\sigma^{}_3\sigma^{}_4}$ is defined by
\be
\bl
W^{\bk\bq,\xi\xi'}_{\sigma^{}_1\sigma^{}_2\sigma^{}_3\sigma^{}_4}=
[\hat{1}+\xi 
\;\!
\hat{\bg}^{}_{\bk}\cdot{\boldsymbol\sigma}]^{}_{\sigma_1\sigma_2}
[\hat{1}+\xi' \hat{\bg}^{}_{\bk}\cdot{\boldsymbol\sigma}]^{}_{\sigma_3\sigma_4}.
\el
\ee
Subsequently, we incorporate the impact of the on-site Hubbard interaction through perturbative renormalization of the initial susceptibility. 
Within the RPA framework, the components of the modified spin susceptibility matrix are determined by the ensuing Dyson equation:
%
%
\begin{align}
	\begin{aligned}
	\Big[\hat{\chi}^{}_{{\rm Ret}}&(\bq,\omega)\Big]^{\sigma_3\sigma_4}_{\sigma_1\sigma_2}
	=
	\Big[\hat{\chi}^{(0)}_{{\rm Ret}}(\bq,\omega)\Big]^{\sigma_3\sigma_4}_{\sigma_1\sigma_2}
	\\
	&+\sum^{}_{\lbrace \alpha_i \rbrace}
	\Big[\hat{\chi}^{(0)}_{{\rm Ret}}(\bq,\omega)\Big]^{\alpha_1\alpha_2}_{\sigma_1\sigma_2}
	\hat{U}^{\alpha_3\alpha_4}_{\alpha_1\alpha_2}
	\Big[\hat{\chi}^{(0)}_{{\rm Ret}}(\bq,\omega)\Big]^{\sigma_3\sigma_4}_{\alpha_3\alpha_4},
	\label{Eq:Dyson}
	\end{aligned}
\end{align}
%
where $\alpha_i$ exhibits the spin indices related to the internal lines in the Feynman diagrams.
Additionally, 
$\hat{U}$ represents the matrix of bare electron-electron interactions in spin space. 
Its non-zero elements are $\hat{U}^{\uparrow\uparrow}_{\downarrow\downarrow}=\hat{U}^{\downarrow\downarrow}_{\uparrow\uparrow}=-U$, and $\hat{U}^{\uparrow\downarrow}_{\downarrow\uparrow}=\hat{U}^{\downarrow\uparrow}_{\uparrow\downarrow}=+U$.
It is crucial to reiterate that in the presence of ASOC, spin-flip scattering occurs. 
Consequently, both the bubble (screening) and ladder (exchange) diagrams must be aggregated to derive the RPA spin susceptibility.
With the spin susceptibilities acquired, we are now poised to investigate the pairing of electrons and the symmetries inherent in the superconducting gap function.
In the Cooper pairs' channel, we examine the interaction between a pair of electrons characterized by momenta and spins $(\bk,\sigma_1)$ and $(-\bk,\sigma_2)$, which are scattered to the states $(\bk',\sigma_3)$, and $(-\bk',\sigma_4)$, respectively.
This process can be modelled by the following Hamiltonian
%
\begin{equation}
	{\cal H}^{\rm RPA}_{\rm Int}=
	\frac{1}{N}\sum_{\bk\bk',\lbrace\sigma_i\rbrace}
	\hat{\Gamma}^{\bk,\bk^\prime}_{\sigma_1\sigma_2\sigma_3\sigma_4}
	c^{\dagger}_{\bk'\sigma^{}_3}
	c^{\dagger}_{-\bk'\sigma^{}_4}
	c^{}_{-\bk\sigma^{}_2}
	c^{}_{\bk\sigma^{}_1},
	\label{Eq:Pair_Interaction}%
\end{equation}
%
where, $\hat{\Gamma}^{\sigma_3\sigma_4}_{\sigma_1\sigma_2}
(\bk,\bk^\prime)$ is the vertex function including the contributions of longitudinal (screening) and
transverse (exchange) interactions.
Separating the vertex function into the singlet (momentum symmetric) and
triplet (momentum antisymmetric) channels leads to
%
\begin{align}
	\begin{aligned}
	\hat{\Gamma}^{{\rm Sing};\bk,\bk^\prime}_{\sigma_1\sigma_2\sigma_3\sigma_4}
 &=
 \\
 &\hspace{-0.5cm}
 \frac{1}{2}\Big[
	\hat{U}
	+
	\frac{3}{2}
	\hat{U}
	\hat{\chi}(\bk-\bk^{\prime},0)
	\hat{U}
	+
	\frac{3}{2}
	\hat{U}
	\hat{\chi}(\bk+\bk^{\prime},0)
	\hat{U}
	\Big];
	\\
	\hat{\Gamma}^{{\rm Trip};\bk,\bk^\prime}_{\sigma_1\sigma_2\sigma_3\sigma_4}&=
	-\frac{1}{2}\Big[
	\hat{U}
	\hat{\chi}(\bk-\bk^{\prime},0)
	\hat{U}
	-
	\hat{U}
	\hat{\chi}(\bk+\bk^{\prime},0)
	\hat{U}
	\Big].
	\end{aligned}
	\label{Eq:Vertex_Sing_Trip}%
\end{align}
%
It's important to note that we solely focus on the static form of the vertex function, specifically when the frequency $(\omega)$ equals zero. 
Any frequency-dependent characteristics are disregarded in our analysis.
By exclusively examining intraband Cooper pairings within the band basis, we can employ the following transformation to determine the effective pairing interaction between two electrons situated on the Fermi surface
%
\begin{equation}
	V^{\rm Sing/Trip}_{\bk,\bk^\prime;\xi\xi^\prime}=
	\sum_{\lbrace\sigma_i\rbrace}
	\hat{\Gamma}^{{\rm Sing/Trip};\bk,\bk^\prime}_{\sigma_1\sigma_2\sigma_3\sigma_4}
	\Lambda^{\xi,*}_{\bk'\sigma^{}_3}
	\Lambda^{\xi,*}_{-\bk'\sigma^{}_4}
	\Lambda^{\xi^\prime}_{-\bk\sigma^{}_2}
	\Lambda^{\xi^\prime}_{\bk\sigma^{}_1},
	\label{Eq:Effective_Interaction_Sing_Trip}%
\end{equation}
%
wherein $\Lambda^{\xi}_{\bk,\sigma}=\langle\bk,\sigma|\bk,\xi\rangle$ connects the states $|\bk,\sigma\rangle$, and $|\bk,\xi\rangle$ in spin and band bases, respectively.
Using BCS theory of superconductivity, the self-consistent equation of the order parameter in both singlet and triplet channels are given by
%
\begin{equation}
	\Delta^{}_{\bk,\xi}=
	-\frac{1}{N}
	\sum^{}_{\bk,\xi^{\prime}}
	V^{\rm Sing/Trip}_{\bk,\bk^\prime;\xi\xi^\prime}
	\frac{\Delta^{}_{\bk^\prime,\xi^\prime}}
	{2E^{}_{\bk^\prime,\xi^{\prime}}}
	\tanh\Big(
	\frac{2E^{}_{\bk^\prime,\xi^{\prime}}}{2T}
	\Big).
	\label{Eq:BCS_equation}%
\end{equation}
%
Here, $E^{}_{\bk,\xi}=\sqrt{\varepsilon^{2}_{\bk,\xi}+\Delta^{2}_{\bk,\xi}}$ represents the dispersion of superconducting quasiparticles.
Linearizing the quasiparticle dispersion near the superconducting critical temperature 
results in $E^{}_{\bk,\xi}=|\varepsilon^{}_{\bk,\xi}|$.
Therefore, Eq.~(\ref{Eq:BCS_equation}) is converted to an eigenvalue problem as
%
\begin{equation}
 \lambda^{}_o\Delta^{}_{\bk,\xi}=
	-\frac{1}{(2\pi)^2}
	\sum^{}_{\xi^{\prime}}
	\oint^{}_{\rm FS}
	\frac{dk^\prime_{|\!|}}{v^{\rm F}_{\bk^\prime,\xi^\prime}}
	V^{\rm Sing/Trip}_{\bk,\bk^\prime;\xi\xi^\prime}
	\Delta^{}_{\bk^\prime,\xi^\prime},
	\label{Eq:BCS_Eigenvalue}%
\end{equation}
with $dk^\prime_{\parallel}$, and $v^{\rm F}_{\bk,\xi}=|{\bf\nabla}\varepsilon^{}_{\bk,\xi}|$ show the differential of  momentum tangent to the Fermi surface and the Fermi velocity, respectively.
In Eq.~(\ref{Eq:BCS_Eigenvalue}), the largest value of $\lambda^{}_o$ indicates the dominant superconducting order in singlet and triplet channels and determine its nodal structure.
%

\subsection{Two-photon density matrix}
We aim to learn more about the process of two-photon generation resulting from pair recombination.
To achieve this goal, we need to calculate the two-photon density matrix using second-order time-dependent perturbation theory. 
This calculation will be performed considering the various symmetries associated with the superconducting scenarios within the system.
The elements of the two-photon density matrix are expressed as:
\begin{equation}
	\bl
 \rho^{p^{}_{1}p^{}_{2}}_{p^{}_{3}p^{}_{4}}
	(\bq^{}_{1},\bq^{}_{2};\vartheta)
	=
	\langle\Psi^{}_{t}(\vartheta)|
	a^{\dagger}_{q^{}_{1},p^{}_{1}}
	a^{\dagger}_{q^{}_{2},p^{}_{2}}
	a^{}_{q^{}_{1},p^{}_{3}}
	a^{}_{q^{}_{2},p^{}_{4}}
	|\Psi_{t}(\vartheta)\rangle.
	\label{Eq:Rho}%
 \el
\end{equation}
%
%
Energy conservation dictates that $\omega^{}_{\bq^{}_{1}}\!+\omega^{}_{\bq^{}_{2}}=2E^{}_{g}$, 
where 
$2E^{}_{g}$
represents the minimum  band gap between 
an electron state with dispersion $\varepsilon^{}_{\bk,\xi}$ and its counterpart hole state of energy $-\varepsilon^{}_{\bk,\xi}$; 
refer to the schematic plot in Fig.~\ref{Fig:Fig2}(g).

By utilizing Wick’s theorem, the expectation values of $\rho^{p^{}_{1}p^{}_{2}}_{p^{}_{3}p^{}_{4}}(\bq^{}_{1},\bq^{}_{2};\vartheta)$ provide insights into the contributions arising from the recombination of Cooper pairs in the generation of two entangled photons.
In addition,
%
\begin{equation}
	|\Psi_{t}(\vartheta)\rangle=
	\int^{t}_{-\infty}\!\!\!\!dt^{}_{1}
	\int^{t^{}_{1}}_{-\infty}\!\!\!\!dt^{}_{2}
	{\cal H}^{}_{\rm int}(t^{}_{1})
	{\cal H}^{}_{\rm int}(t^{}_{2})
	|\Psi^{}_{0}\rangle,
	\label{Eq:Psi_t}%
\end{equation}
%
depicts the state of the second-order system at time $t$ in the interaction picture, in which ${\cal H}^{}_{\rm int}(t)\!=\!e^{\mi {\cal H}^{}_{0}t}{\cal H}^{}_{\rm int}e^{-\mi {\cal H}^{}_{0}t}$, where ${\cal H}^{}_{\rm int}(t)$ denotes the interacting potential at time $t$.
Furthermore, $$|\Psi^{}_{0}\rangle=|0\rangle|{\rm FS}\rangle|{\rm BCS}\rangle$$ represents the system's initial state, with $|0\rangle$, $|{\rm FS}\rangle$, and $|{\rm BCS}\rangle$ indicating the vacuum state of photons, the Fermi sea of the heavy hole band, and the BCS wave-function of the superconducting electrons in the conduction band, respectively.
Using Eqs.~(\ref{Eq:Rho}) and (\ref{Eq:Psi_t}), performing the time integrals, and summing over the internal degrees of freedom, we obtain the $\vartheta$-resolved emission rate, given by:
\begin{equation}
	\bar{\varrho^{}_{\vartheta}}(\bq^{}_{1},\bq^{}_{2})=
	\frac{1}{2N}
	\sum^{}_{\bk,\xi}
    \zeta^{\bk,\xi,\vartheta}_{\bq^{}_{1},\bq^{}_{2}}(\psi^{}_{\bk},\bd^{}_{\bk})
	{\cal M}^{}_{\bk,\xi}(\psi^{}_{\bk},\bd^{}_{\bk}),
	\label{Eq:Emission_Rate}%
\end{equation}
in which 
the weight factor $\zeta^{\bk,\xi,\vartheta}_{\bq^{}_{1},\bq^{}_{2}}(\psi^{}_{\bk},\bd^{}_{\bk})$ encapsulates the rate of photon pair generation resulting from the recombination of Cooper pairs, accounting for the specific contributions of Rashba and Dresselhaus SOCs, 
which is given by
%
%
\begin{align}
\begin{aligned}
\zeta^{\bk,\xi,\vartheta}_{\bq^{}_{1},\bq^{}_{2}}
&
(\psi^{}_{\bk},\bd^{}_{\bk})
\!=\!
\pi 
|B^{}_{\bk,\bq^{}_{1}}|^2
|B^{}_{\bk,\bq^{}_{2}}|^2
\Big\vert 
\frac{\Delta^{}_{\bk,\xi}}{E^{}_{\bk,\xi}}
\Big\vert^2
[n^{}_{F}(-E^{}_{\bk,\xi})]^2
\times
\\
&\Bigg[
\Bigg(
\frac{
[n_{F}(E^{}_{\bk,\xi})]^{2}
}{
(E^{}_{\bk,\xi}-\omega^{}_{\bq_1}+\varepsilon_{\bk,\xi})^2
}
+
\frac{
[1-n^{}_{F}(E^{}_{\bk,\xi})]^2
}{
(E^{}_{\bk,\xi}+\omega^{}_{\bq_1}-\varepsilon_{\bk,\xi})^2
}
\\
&
+
\frac{2
n^{}_{F}(E^{}_{\bk,\xi})
[1-n_{F}(E^{}_{\bk,\xi})]
}{
(E^{}_{\bk,\xi}-\omega^{}_{\bq_1}+\varepsilon_{\bk,\xi})
(E^{}_{\bk,\xi}+\omega^{}_{\bq_1}-\varepsilon_{\bk,\xi})
}
\\
&+
\frac{
[n_{F}(E^{}_{\bk,\xi})]^{2}
}{
(E^{}_{\bk,\xi}-\omega^{}_{\bq_1}+\varepsilon_{\bk,\xi})
(E^{}_{\bk,\xi}-\omega^{}_{\bq_2}+\varepsilon_{\bk,\xi})
}
\\
&
+
\frac{
n^{}_{F}(E^{}_{\bk,\xi})
[1-n^{}_{F}(E^{}_{\bk,\xi})]
}{
(E^{}_{\bk,\xi}+\omega^{}_{\bq_1}-\varepsilon_{\bk,\xi})
(E^{}_{\bk,\xi}-\omega^{}_{\bq_2}+\varepsilon_{\bk,\xi})
}
\\
&+
\frac{
n^{}_{F}(E^{}_{\bk,\xi})
[1-n^{}_{F}(E^{}_{\bk,\xi})]
}{
(E^{}_{\bk,\xi}-\omega^{}_{\bq_1}+ \varepsilon_{\bk,\xi})
(E^{}_{\bk,\xi}+ \omega^{}_{\bq_2}-\varepsilon_{\bk,\xi})
}
\\
&
+
\frac{
[1-n^{}_{F}(E^{}_{\bk,\xi})]^2
}{
(E^{}_{\bk,\xi}+\omega^{}_{\bq_1}-\varepsilon_{\bk,\xi})
(E^{}_{\bk,\xi}+\omega^{}_{\bq_2}-\varepsilon_{\bk,\xi})
}
\Bigg)
\\
&
+
(\bq^{}_{1}\longleftrightarrow\bq^{}_{2})
\Bigg]
%
%
\delta(\omega^{}_{\bq^{}_1}+\omega^{}_{\bq^{}_2}-2\varepsilon^{}_{\bk,\xi}).
\end{aligned}
\label{Eq:Zeta}
\end{align}
%
%

%
%
%
\begin{figure}[t]
	\includegraphics[width=\linewidth]{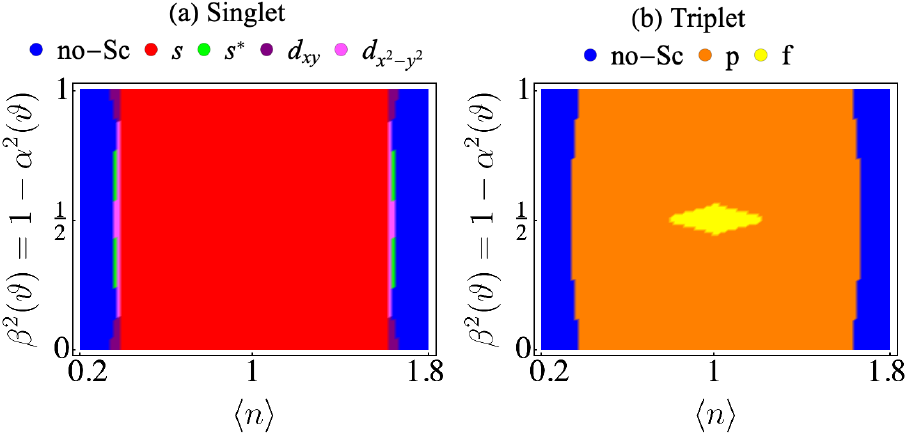}
	\caption{
	The phase diagram of superconducting order parameter for (a) singlet, and (b) triplet channels with respect to electron filling $\langle n \rangle$, and relative amplitude of Rashba and Dresselhaus SOCs at $U=0.8 t$. See Ref.~\cite{Biderang_arXiv_2019} for more discussion. 
	}
	\label{Fig:Fig3}
\end{figure}
%

%
%
%
%
In the limit of $\lambda\rightarrow 0$, Eq.~(\ref{Eq:Emission_Rate}) converges to the expression reported in Ref.~\cite{Gordon_PRB_2018}.
The basis for the two-photon density matrix is describes by the circular right- or left-handed polarization and is given by $(|\!L\!L\rangle,|\!L\!R\rangle,|\!R\!L\rangle,|\!R\!R\rangle)$.
Under this basis, the matrix ${\cal M}^{}_{\bk,\xi}(\psi^{}_{\bk},\bd^{}_{\bk})$ is defined as:
\begin{equation}
\bl
&
		{\cal M}^{}_{\bk,\xi}(\psi^{}_{\bk},\bd^{}_{\bk})
		=
            \begin{pmatrix}
			\Upsilon_{\bk,\xi}  & 0 & 0 & 0 \\
			0 & \eta_{\bk,\xi}
			& \eta_{\bk,\xi} & 0 \\
			0 & \eta_{\bk,\xi}
			& \eta_{\bk,\xi} & 0 \\
			0 & 0 & 0 &  
			\Upsilon_{\bk,\xi} \\
		\end{pmatrix},
		\label{Eq:Emission_Rate_Matrix}
 \el
 \end{equation}
where 
$$
\eta_{\bk,\xi}=
{\cal A}^{2}_{\bk,\xi}+{\cal B}^{2}_{\bk,\xi} \cos^{2}_{}\!\varsigma^{}_{\bk};
\quad
\Upsilon_{\bk,\xi}
=
2{\cal B}^{2}_{\bk,\xi} \sin^{2}_{}\!\varsigma^{}_{\bk}
$$
with 
 the fractions of pairings in singlet 
${\cal A}^{}_{\bk,\xi}=
\psi^{}_{\bk}/|\!|\hat{\Delta}^{}_{\bk}|\!|$,
and triplet channels
${\cal B}^{}_{\bk,\xi} = \xi
|\bd^{}_{\bk}|
/
|\!|\hat{\Delta}^{}_{\bk}|\!|$.
Here 
$$
|\!|
\hat{\Delta}^{}_{\bk}
|\!| =
\Big[
\prod^{}_{\xi}\Delta^{}_{\bk,\xi}
\Big]^{\frac{1}{2}}
=\sqrt{\psi^{2}_{\bk}+|\bd^{}_{\bk,\xi}|^2},
$$
and
$\varsigma^{}_{\bk}\!\!=\!
\arccos(\hat{\bp}\cdot\bd^{}_{\bk}/|\bd^{}_{\bk}|)$
represents the angle between $\bd^{}_{\bk}$ and the polarization axis of photons $\hat{\bp}=(\hat{x}\cos\phi+\hat{y}\sin\phi)\sin\theta+\hat{z}\cos\theta$.
The fractions of pairings satisfy the  property
${\cal A}^{2}_{\bk,\xi}+{\cal B}^{2}_{\bk,\xi}=1$.
\\

%
%
\begin{figure}[t]
	\includegraphics[width=0.99\linewidth]{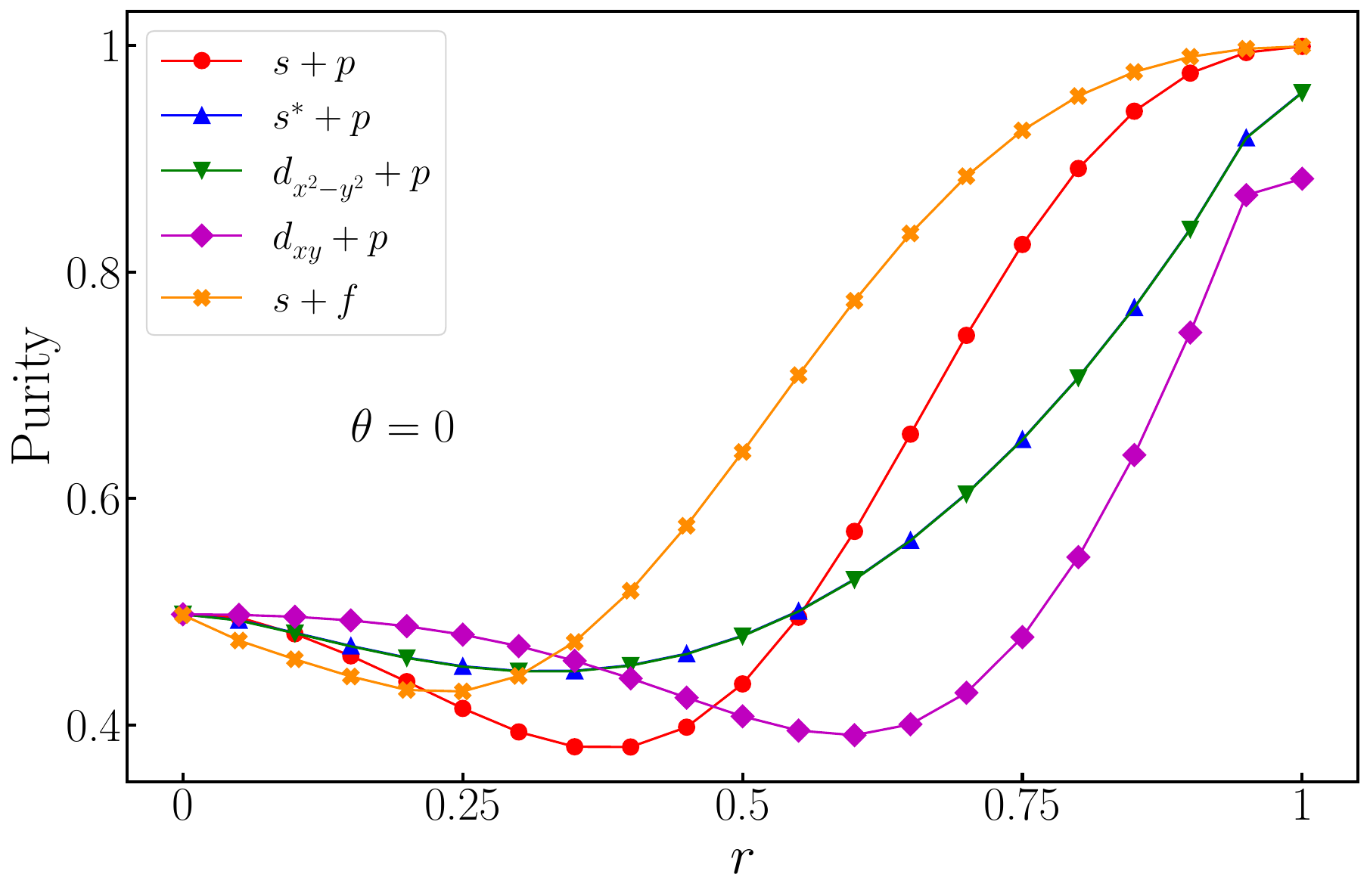}
	\caption{
	The purity of photon pairs, $\Gamma$, for various combinations of superconducting gap functions, considering different amplitudes of singlet and triplet channels for 
$\theta^{}_{}=0$. 
	Notably, $r=0$ and $r=1$ represent instances of pure triplet and singlet pairings, respectively.
	}
	\label{Fig:Fig4}
\end{figure}
%

%
\begin{figure}[t]
	\includegraphics[width=0.99\linewidth]{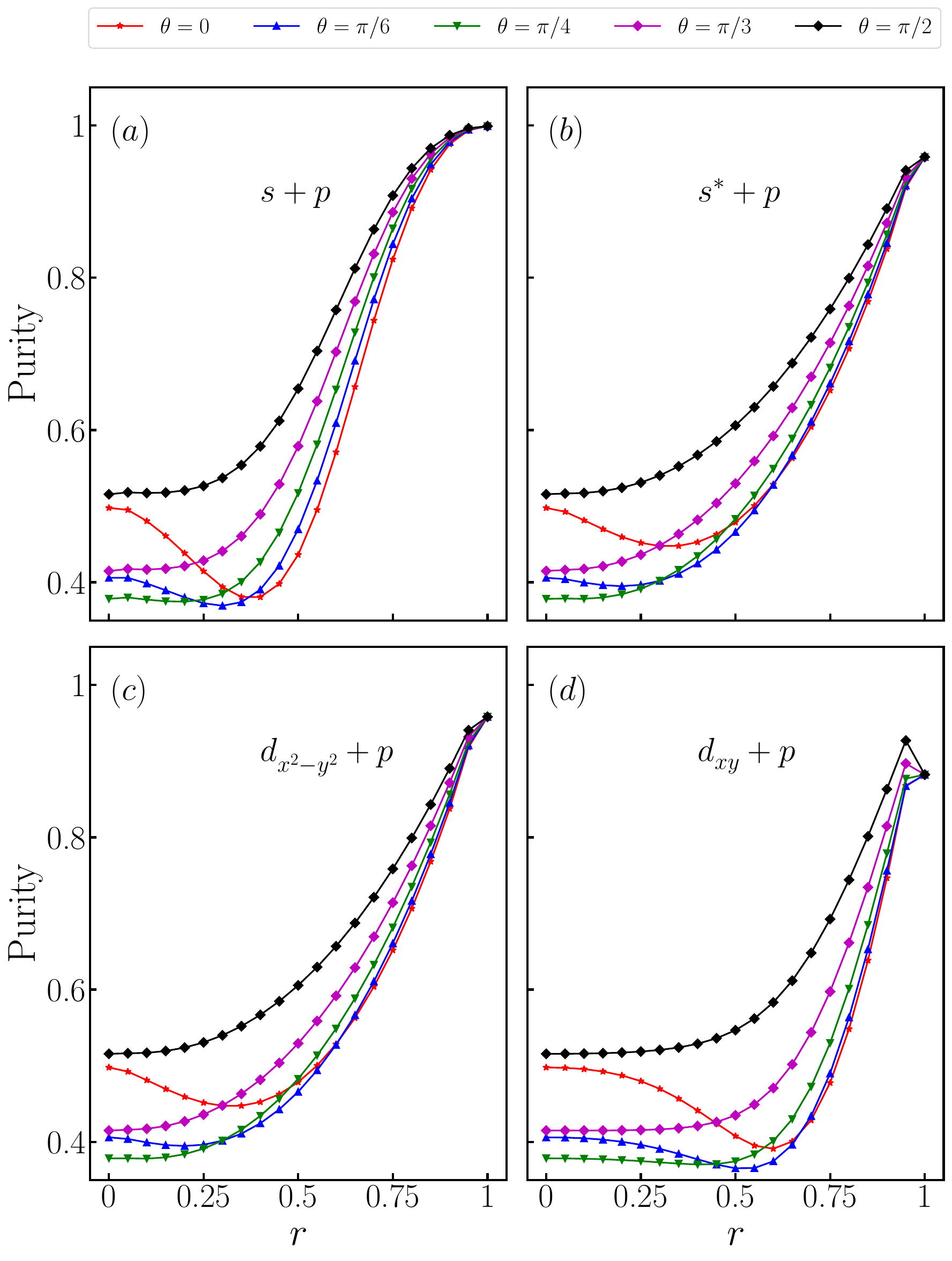}
	\caption{
		The purity of generated entangled photon pairs, $\Gamma$, with respect to the amplitudes of singlet and triplet pairings for the various gap functions. 
        Each curve reports at different values $\theta$, by setting  $\phi=0$.
		}
	\label{Fig:Fig5}
\end{figure}
%

Excluding the contribution of the weight factor $\zeta^{\bk,\xi,\vartheta}_{\bq^{}_{1},\bq^{}_{2}}(\psi^{}_{\bk},\bd^{}_{\bk})$, the purity of the two-photon states is obtained by 
%
\begin{equation}
\bl
\Gamma
&=
\frac{1}{2N}
\sum^{}_{\bk,\xi}{\rm Tr}[{\cal M}^{}_{\bk,\xi}(\psi^{}_{\bk},\bd^{}_{\bk})]^2.
\\
&=
\frac{1}{2N}
\sum^{}_{\bk,\xi}
\Big[
{\cal B}^{4}_{\bk,\xi}
\sin^{4}_{}\!\varsigma^{}_{\bk}
+
2
\Big(
{\cal A}^{2}_{\bk,\xi}
+
{\cal B}^{2}_{\bk,\xi}
\cos^{2}_{}\!\varsigma^{}_{\bk}
\Big)^2
\Big].
\el
\end{equation}
%
Here, it is important to note that since we have not considered the influence of Rashba and Dresselhaus contributions on the structure of the superconducting gap function, the matrix ${\cal M}^{}_{\bk,\xi}(\psi^{}_{\bk},\bd^{}_{\bk})$ remains $\vartheta$-independent.
However, the emission rate matrix 
$\bar{\varrho^{}_{\vartheta}}$ is obviously $\vartheta$-dependent.
%
%

%
\begin{figure}[t]
	\includegraphics[width=0.99\linewidth]{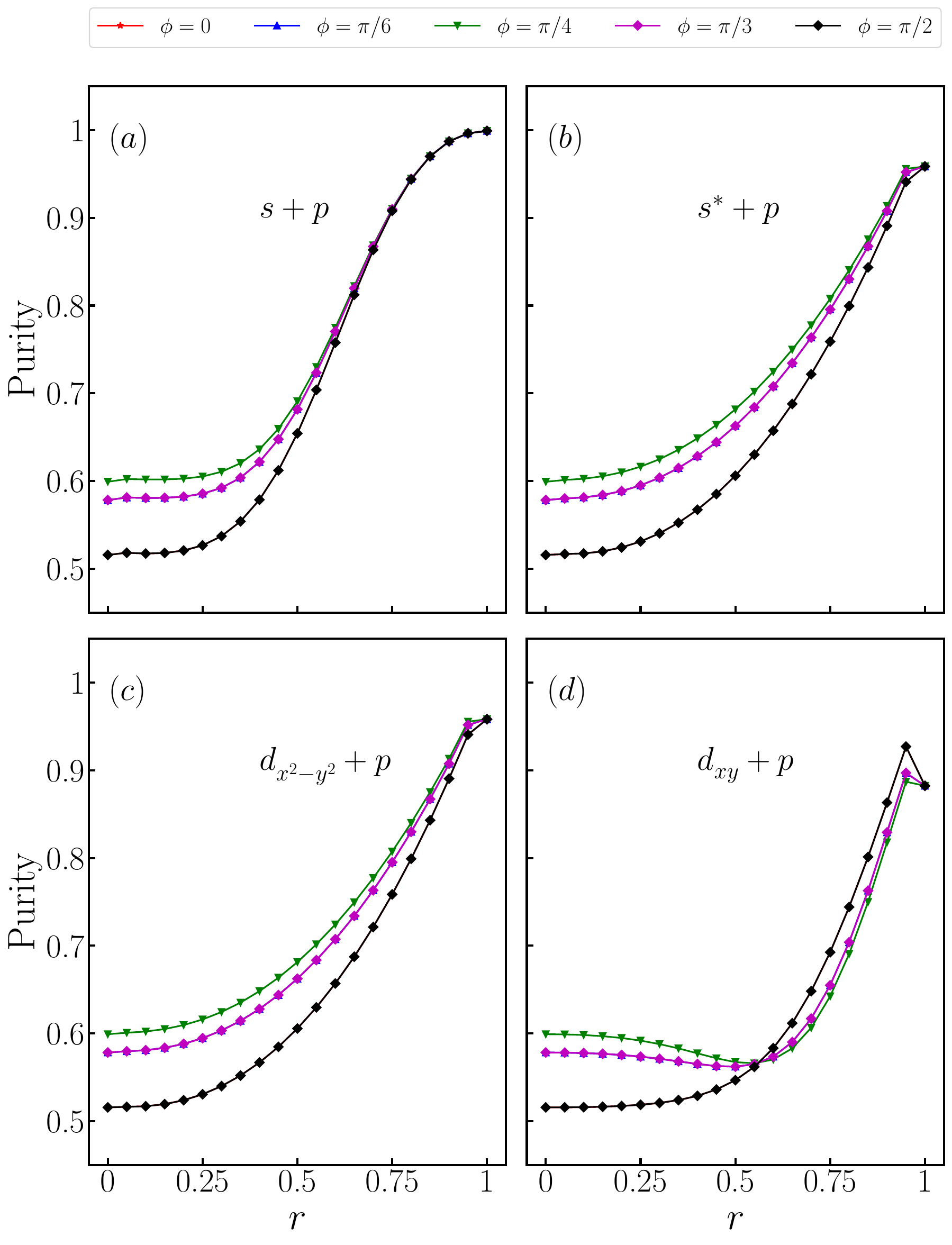}
	\caption{
		Variation in the purity of entangled photons concerning the parameter $r$ across different values of the azimuthal angle $\phi$, at $\theta=\pi/2$. 
		This scenario corresponds to the polarization axis of photons lying in the $xy$-plane. 
       Note that the curves corresponding to $\phi=0$ and $\phi=\pi/2$ exhibit identical patterns, as do those for $\phi=\pi/6$ and $\phi=\pi/3$.
	}
	\label{Fig:Fig6}
\end{figure}
%

%
\begin{figure}[t]
	\includegraphics[width=\linewidth]{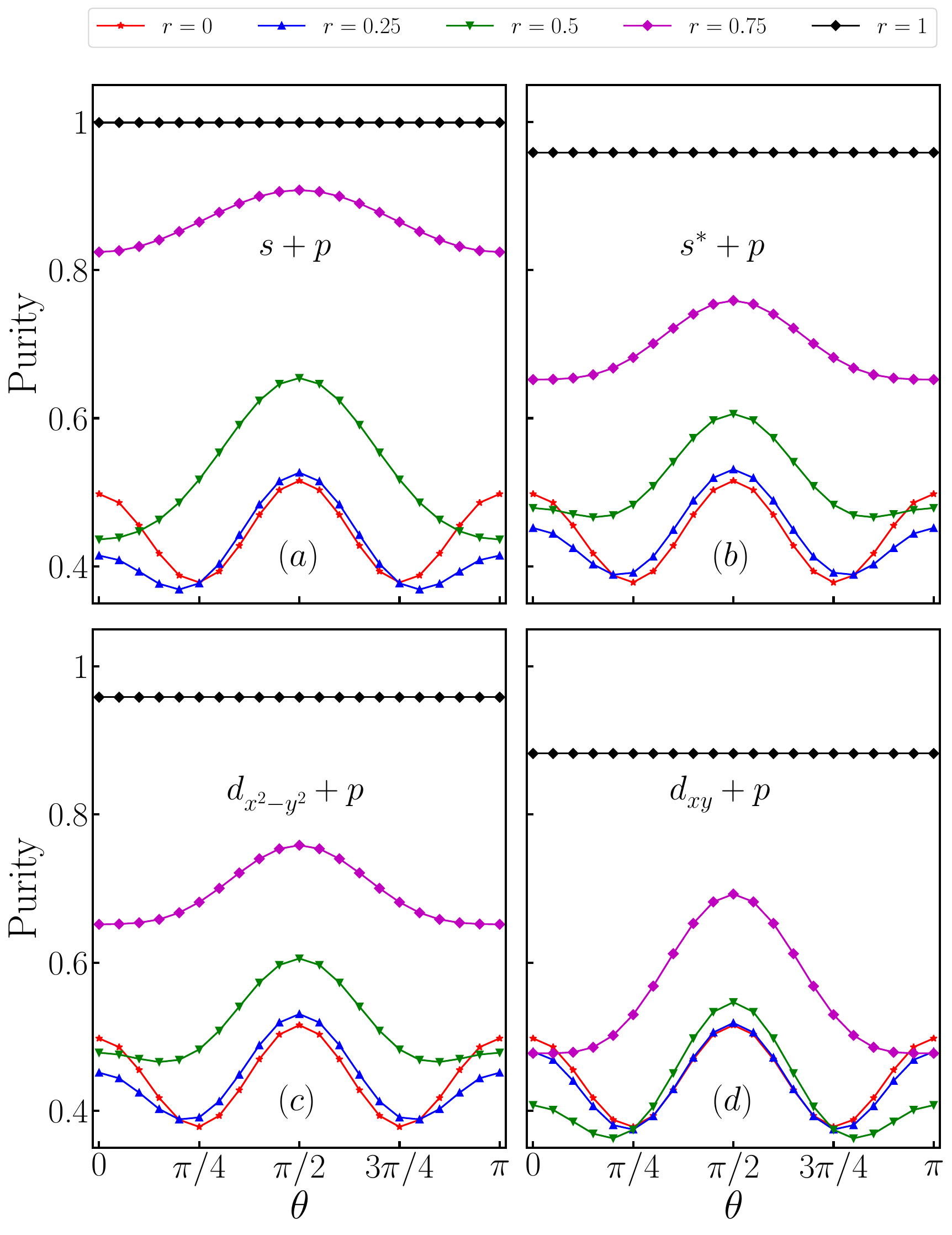}
	\caption{
		The impact of variations in the polar angle $\theta$ on the purity of entangled photon pairs. 
		The figure presents diverse contributions of singlet and triplet pairings, specifically at $\phi=0$, displaying various ratios of singlet and triplet amplitudes.
	}
	\label{Fig:Fig7}
\end{figure}
%

\begin{figure}[t]
	\includegraphics[width=\linewidth]{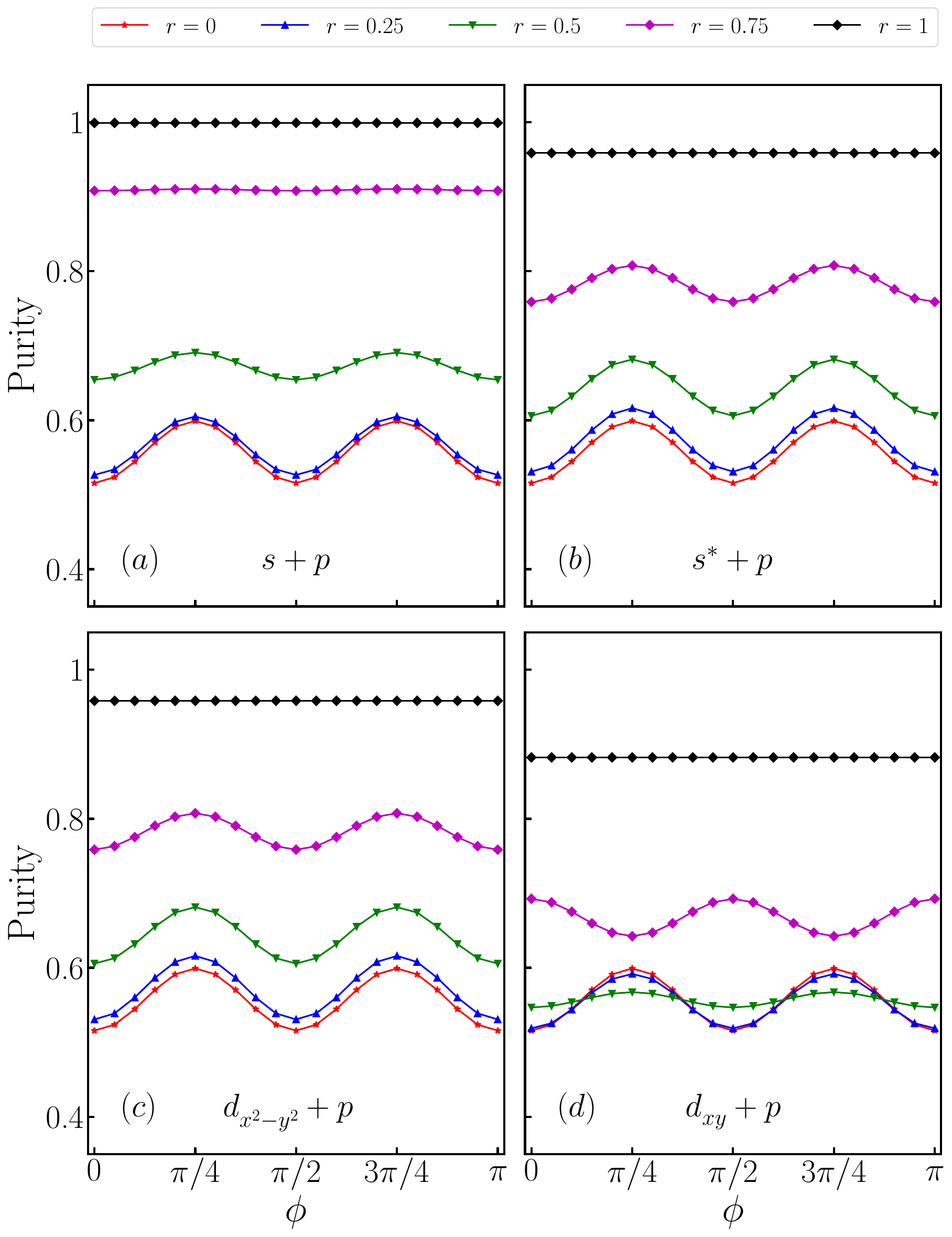}
	\caption{
		Demonstrating the impact of variations in the azimuthal angle $\phi$ on the purity of entangled photon pairs. 
		The figure showcases diverse contributions of singlet and triplet pairings, specifically at $\theta=\pi/2$, where the $\hat{\bp}$-vector resides in the $xy$-plane, presenting various ratios of singlet and triplet amplitudes.
	}
	\label{Fig:Fig8}
\end{figure}
%
%
\begin{figure}[t]
	\includegraphics[width=\linewidth]{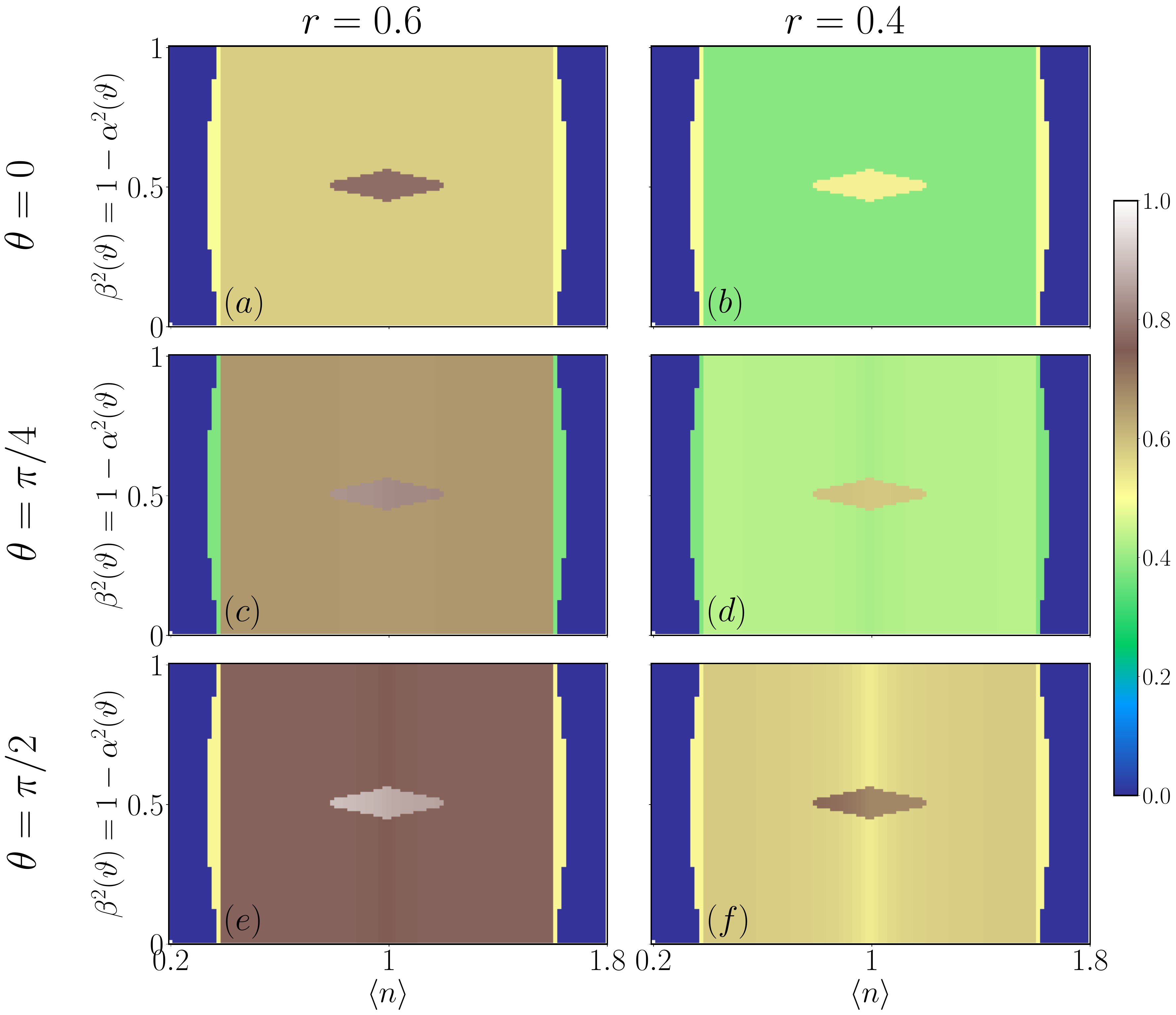}
	\caption{
		Purity of generated entangled photons regarding the concentration of charge carriers and relative amplitudes of Rashba/Dresselhaus SOCs at $\phi=0$.
		The left and right columns represent the singlet ($r=0.6$), and triplet ($r=0.4$) dominant pairings.
	}
	\label{Fig:Fig9}
\end{figure}
%

%
\begin{figure}[t]
	\includegraphics[width=\linewidth]{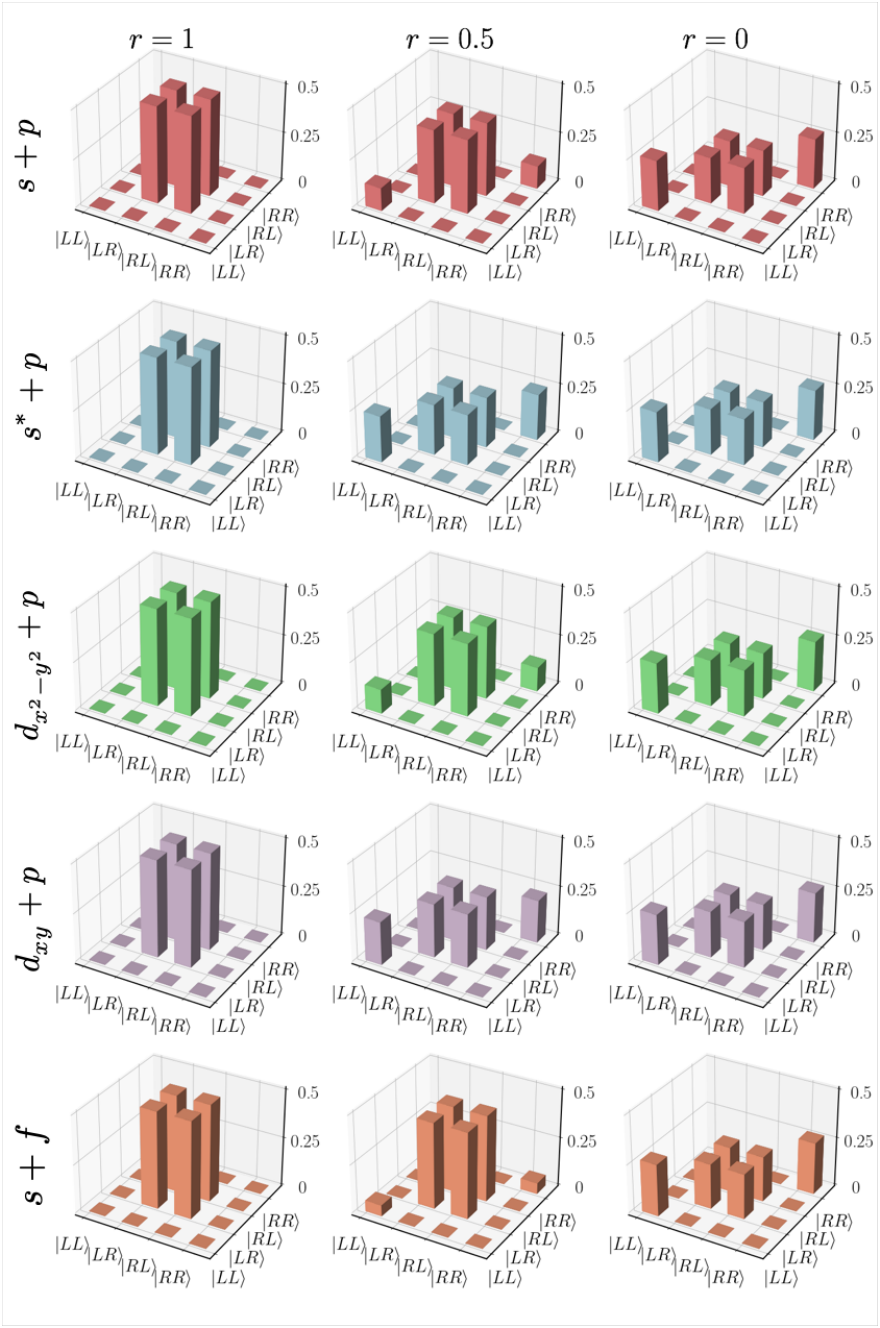}
	\caption{
		Density matrix of polarized two-photon states generated through the recombination of Cooper pairs for $\theta=\pi/2$, and $\phi=0$. 
		The  left, middle, and right panels  illustrate the outcomes for pure singlet ($r=1$), an equal mix of singlet and triplet ($r=0.5$), and pure triplet ($r=0$) pairings, respectively. Within each column, the rows correspond to different gap functions: $s+p$, $s^{*}+p$, $d^{}_{x^2-y^2}+p$,  $d^{}_{xy}+p$, and $s+f$. 
	}
	\label{Fig:Fig10}
\end{figure}
%

\section{Results and Discussion}
\label{Sec:Results}
We initially study the impact of changes in the concentration of charge carriers and the relative amplitude of Rashba/Dresselhaus spin-orbit couplings on the structure of the superconducting gap within weak coupling theory.
Fig.~\ref{Fig:Fig3} illustrates the superconducting phase diagrams derived from Eq.~(\ref{Eq:BCS_Eigenvalue}) concerning the charge carrier filling and the relative amplitudes of antisymmetric Rashba/Dresselhaus SOCs, in both singlet and triplet channels.
In Fig.~\ref{Fig:Fig3}(a), the superconducting gap texture in the singlet channel is presented. %
It is apparent that over large regions of filling and relative contributions of Rashba and Dresselhaus SOCs, $s$-wave pairing predominates over other singlet pairings.
Only in a very narrow region 
far from half-filling the other singlet pairings of higher orbital angular momenta appear.
Fig.~\ref{Fig:Fig3}(b) displays the momentum dependence of the triplet component of Cooper pairing. It reveals that except for a small area around half-filling and equal contributions of Rashba/Dresselhaus SOCs, where $f$-wave pairing is stable, the triplet pairing exhibits a $p$-wave texture in other regions.
It should be noted that Dresselhaus spin-orbit coupling arises as a bulk effect due to bulk inversion asymmetry (BIA), typically exhibiting a cubic momentum dependence in bulk materials. 
However, during crystal growth processes involving dimensional confinement, this cubic momentum dependence can transform into special linear dependencies. 
Consequently, for certain materials oriented along specific crystal growth directions, the amplitude of Dresselhaus SOC remains constant.
In contrast, Rashba SOC stems from structural inversion asymmetry (SIA) and can be induced and manipulated using external electric fields or gate voltages. 
Hence, the relative amplitudes of Rashba and Dresselhaus SOCs are adjustable via external electric field modulation.
These distinctions highlight the different origins and controllability of Rashba and Dresselhaus SOCs, underscoring their potential for manipulation in electronic and spintronic devices.
%
%

%
\begin{figure}[b]
\includegraphics[width=0.9\linewidth]{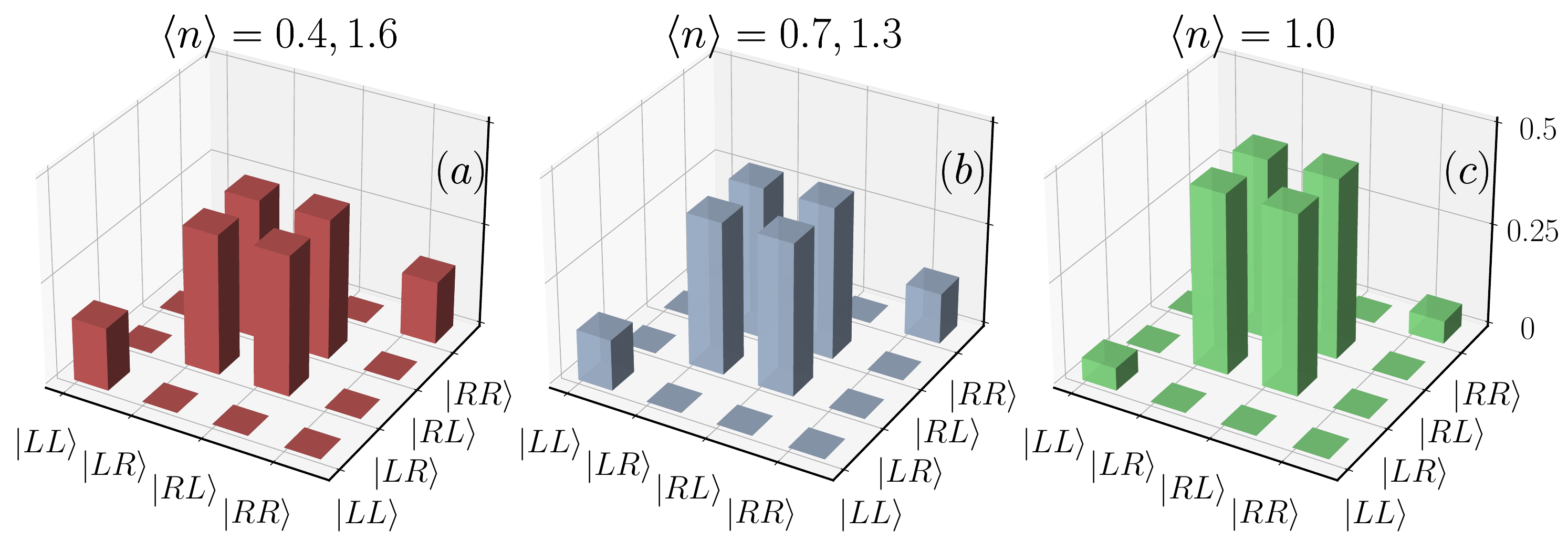}
	\caption{
		Influence of hole/electron-filling on the density matrix of polarized two-photon states at $\vartheta=\pi/4$, $\theta=\pi/2$, and $\phi=0$ for the same contributions of singlet and triplet channels ($r=0.5$). 
		Due to the intrinsic electron-hole symmetry of band structure, the corresponding electron-doped cases follow the same pattern with the electron-doped one.
	}
	\label{Fig:Fig11}
\end{figure}
%

%
Our investigation now focuses on assessing the purity of the polarization state of the two photons, considering various scenarios dictated by the interplay of distinct physical parameters involved in the process.
For a specific scenario where the photon's polarization axis is $z$-direction ($\theta=0$), the purity of the two-photon state, $\Gamma$, is depicted in Fig.~\ref{Fig:Fig4}, concerning the diverse admixture of singlet and triplet pairings across various superconducting gap functions. 
When $r=1$, representing pure triplet $p$-wave pairing, the purity reaches its minimum value. Intriguingly, the addition of singlet pairing alongside dominant triplet Cooper pairs leads to a reduction in purity. However, around $r=0.5$, where singlet and triplet components of the gap function are of comparable magnitudes, purity begins to increase.
At $r=1$, where the superconducting gap function demonstrates a pure singlet texture, the purities achieve their maximum values. Notably, conventional singlet $s$-wave pairing exhibits the highest achievable purity. Consequently, conventional $s$-wave superconductors emerge as the most suitable candidates for generating entangled photons via Cooper pair recombination.
Fig.~\ref{Fig:Fig5} showcases the purity of polarized photons concerning the parameter $r$ for various
$\theta$ by considering $\phi=0$. For the case of $\theta=\pi/2$, where the polarization vector lies in the $xy$-plane, entangled photon purity maximizes across different superconducting pairing symmetries.
Specifically, for pure triplet $p$-wave pairing ($r=0$), maximum purity is attained when both $\hat{\bp}$ and $\bd_\bk$ vectors lie in the same plane.
Furthermore, Fig.~\ref{Fig:Fig6} exhibits the purity versus the singlet/triplet pairing ratio for various values of the azimuthal angle $\phi$ at $\theta=\pi/2$. Notably, for $\phi=\pi/4$, the purity of generated entangled photons reaches its peak.
Additionally, owing to the intrinsic $C^{}_{4v}$ point group symmetry, complementary angles yield identical results.
To assess the impact of the angle $\varsigma^{}_{\bk}$ on the purity behaviour of generated entangled photons,
Figs.~\ref{Fig:Fig7} and ~\ref{Fig:Fig8} depict the behaviour of entangled photon purity concerning changes of $\theta$ and $\phi$ angles across various combinations of singlet/triplet pairings.
It's noticeable that the purity for pure singlet pairing remains constant regardless of these angles. These graphs affirm that the highest purity is achievable only in the case of $s$-wave pairing.
Moreover, within the singlet channel, the $s$- and $d^{}_{xy}$-wave pairings exhibit the maximum and minimum values of purity, respectively.
The $C^{}_{4v}$ symmetry of the crystalline lattice dictates that the purity pattern remains unchanged under a $\phi=\pi/4$ rotation around the $z$ axis.
So far, we have only explored the impact of the superconducting gap texture on the purity of the generated entangled photons.
To investigate the effect of the band structure on the purity of the photons, we conducted calculations involving the purity of the entangled photons with respect to filling and relative amplitudes of SOCs.
Fig.~\ref{Fig:Fig9} exhibits the purity of generated entangled photons with respect to the filling and relative strengths of Rashba/Dresselhaus SOCs based on the superconducting phase diagram depicted in Fig.~\ref{Fig:Fig3}.
It is observed that for every specific value of $r$ and $\theta$, the highest purity is achieved around half-filling, where equal contributions of Rashba-Dresselhaus SOCs prevail.
In this region, the singlet and triplet channels of the superconducting ground state have 
$s$- and $f$-wave couplings, respectively.
These results are entirely consistent with our findings and support our conclusion that singlet-dominant pairings exhibit greater purity compared to triplet-dominant superconductivity. 
Furthermore, it's observed that higher values of $\theta$ correspond to higher purity of generated entangled photons.
Now, we are exploring how various textures of the superconducting ground state in noncentrosymmetric systems impact the purity of entangled photons. This investigation involves analyzing the matrix elements of the $\vartheta$-resolved emission rate.
We present the emission rate $\bar{\varrho}^{}_{\vartheta}(\bq^{}_{1},\bq^{}_{2})$,  in Fig.~\ref{Fig:Fig10}, for various superconducting gap function textures across different singlet-triplet admixture ratios $r$.
The resulting two-photon states exhibit correlations that could lead to the generation of entangled states such as $|\Psi^{}_{\rm ph}\rangle=(|LR\rangle\pm|RL\rangle)/\sqrt{2}$.
Therefore, for the production of entangled two-photon states, the focus should be on attaining higher purity in generating $|LR\rangle$ and $|RL\rangle$ states.
Fig.~\ref{Fig:Fig10} reveals that in the case of pure singlet pairing $(r=1)$, only $|LR\rangle$ and $|RL\rangle$ two-photon states are produced.
However, inducing triplet pairing due to the presence of spin-orbit coupling  results in the generation of $|LL\rangle$ and $|RR\rangle$ states, which do not lead to entangled pairs.
Given that Rashba and/or Dresselhaus SOCs, 
triplet Cooper pairs exist alongside singlet pairs leading to reduced purity in entangled two-photon states.
To achieve higher purity, it is crucial to control the strength of antisymmetric SOC to decrease the amplitude of odd parity spin-triplet superconductivity.
Furthermore, we find that within the parity-mixed superconducting states, the scenario with $s+f$-wave pairings exhibits the most significant occurrence of $|LR\rangle$ and $|RL\rangle$ states.
This configuration is more predisposed to generate entangled two-photon pairs.

Fig.~\ref{Fig:Fig11} shows the influence of  charge carrier filling, on the amplitudes of the different matrix elements of the $\vartheta$-resolved emission rate for $\vartheta=\pi/4$ and $r=0.5$.
This result demonstrates that in the vicinity of half-filling, the amplitudes of $|LR\rangle$ and $|RL\rangle$ states reach their maximum values, indicating a heightened propensity for entanglement. %
Conversely, the populations of $|RR\rangle$ and $|LL\rangle$ states are minimized under similar conditions. 
This intriguing observation not only underscores the enhanced likelihood of entanglement but also suggests a significant role played by the electronic structure near half-filling in promoting entangled states. 
Such insights pave the way for further exploration into the underlying mechanisms governing the emergence and dynamics of entanglement in noncentrosymmetric systems.
%

\section{Summary}
\label{Sec:Summary}
We have conducted an exploration into the generation of entangled two-photon states through Cooper pair recombination within a P-N-S heterostructure. 
The semiconductor components utilized in our investigation adopted a zinc-blend crystal structure, embodying a combination of Rashba and Dresselhaus antisymmetric SOCs.
Our research encompassed an analysis of various types of superconducting gap functions, considering diverse ratios between singlet and triplet pairings.
Additionally, we scrutinized the influence of alterations in the contribution of Rashba/Dresselhaus SOCs alongside electron filling levels on both the purity and prevalence of distinct pairs of polarized two-photon states.
Given the inevitable presence of spin-triplet Cooper pairs in structures and compounds featuring antisymmetric spin-orbit coupling, we consider examining the ramifications of altering the angle 
between the spin-triplet pairing orientation, 
and the polarization axis of photons. 
In accordance with the previous results, our calculations reinforce the notion that pure singlet pairings yield the highest purity levels compared to both mixed and pure triplet pairings. 
Additionally, within the singlet gap functions, we observed that the $s$- and $d^{}_{xy}$-wave pairings exhibit the highest and lowest purity levels, respectively.
In noncentrosymmetric crystals featuring antisymmetric spin-orbit coupling, the induced spin-triplet pairing tends to diminish the population of entangled two-photon states. 
Our findings further revealed that in superconducting ground states with mixed-parity pairings, the purity amplitude of entangled two-photon states reaches its peak at $(\theta,\phi)=(\pi/2,\pi/4)$.
Finally, our  analysis concludes that in the vicinity of half-filling, particularly when the Rashba and Dresselhaus spin-orbit couplings  exhibit identical amplitudes, the superconducting gap structure manifests as a combination of $s+f$-wave pairings. 
Remarkably, under these conditions, both the purity and population of the generated entangled two-photon pairs reach their peak values. 
This notable correlation underscores the critical role played by the electronic configuration near half-filling, coupled with the specific symmetry properties of the spin-orbit couplings, in facilitating the emergence of maximally entangled states. 
These findings offer valuable insights into the intricate interplay between electronic structure, superconducting pairing mechanisms, and the generation of entanglement in noncentrosymmetric systems, thereby advancing our understanding of quantum phenomena in condensed matter physics.
\\
%

	
\section*{Acknowledgments}
M.~B. and A.~A. highly appreciate Hae-Young Kee for the inspiring discussions.
A.~A. acknowledges the financial support from the German Research Foundation within the bilateral NSFC-DFG Project No. ER 463/14-1.
%
%
\bibliography{Main_Refs}
\end{document}